\documentstyle[12pt]{article}
\renewcommand
\baselinestretch{1.3}

\begin{document}

\hfill December 1993

\vspace*{3mm}

\begin{center}

{\LARGE \bf Chiral symmetry breaking in the Nambu-Jona-Lasinio model
in curved spacetime with non-trivial topology}

\vspace{4mm}

\renewcommand
\baselinestretch{0.8}
\medskip

\renewcommand
\baselinestretch{1.4}
{\sc E. Elizalde}\footnote{E-mail: eli @ ebubecm1.bitnet, eli @
zeta.ecm.ub.es} \\ Department E.C.M. and I.F.A.E., Faculty of Physics,
University of  Barcelona, \\ Diagonal 647, 08028 Barcelona, \\
and Center for Advanced Studies, C.S.I.C., Cam\'{\i} de Santa B\`arbara, \\
17300 Blanes, Catalonia, Spain \\
{\sc S. Leseduarte}\footnote{E-mail: lese @ ebubecm1.bitnet} and
{\sc S.D. Odintsov}\footnote{E-mail: odintsov @ ebubecm1.bitnet.
On
leave from: Tomsk Pedagogical Institute, 634041 Tomsk, Russian
Federation.} \\  Department E.C.M., Faculty of Physics,
University of  Barcelona, \\  Diagonal 647, 08028 Barcelona, Catalonia,
Spain \\

\vspace{5mm}

{\bf Abstract}

\end{center}

We discuss the phase structure (in the $1/N$-expansion) of the
Nambu-Jona-Lasinio model in curved spacetime with non-trivial topology
${\cal M}^3 \times {\rm S}^1$. The evaluation of the effective potential
of the composite field $\bar{\psi} \psi$ is presented in the linear
curvature approximation (topology is treated exactly) and in the leading
order of the $1/N$-expansion. The combined influence of topology and
curvature to the phase transitions is investigated. It is shown, in
particular, that at zero curvature and for small radius of the torus
there is a second order phase transition from the chiral symmetric to
the chiral
non-symmetric phase. When the curvature grows and (or) the radius
of
${\rm S}^1$ decreases, then the phase transition is in general of first
order.
The dynamical fermionic mass is also calculated in a number of different
situations.

\vspace{4mm}

\newpage

\section{Introduction}
 The study of composite fermionic fields ---and effects related with
them--- in the very early universe has attracted much attention in the
physical community.
In order to be able to
do such kind of research, one must first develop an effective action
formalism for
composite fields in a general curved spacetime (for an introduction to
this subject, see \cite{1}).
However, even in flat space only  very few models are known which
can be treated analytically, when studying the composite boundstates.
The Nambu-Jona-Lasinio (NJL) model \cite{2}
(for a recent discussion, see \cite{3}) belongs to such interesting
class. This model is usually discussed
in frames of the $1/N$-expansion (see, for example \cite{6}),
which is a very useful scheme
to study the non-perturbative effects. The NJL model may be
considered as an effective field theory for QCD in some region, and it is
also connected with the theory of superconductivity (related models
with quite interesting properties were studied some time ago \cite{es}).
One can explicitly
discuss the dynamical chiral symmetry breaking for the NJL model in the
$1/N$-expansion, where a vacuum condensate $<\bar{\psi}
\psi > \neq 0$ appears and a dynamical fermionic mass is generated.

 Recently \cite{4,5}, a detailed study of the NJL model in curved
spacetime has been started. The one-loop effective potential in the
$1/N$-expansion and in the linear curvature approximation has
been
calculated and the existence of a curvature-induced first-order phase
transition from a chiral symmetric to a chiral non-symmetric phase has
been
shown \cite{5} (for a review on curvature-induced phase transitions for
elementary fields in GUT's see, for example \cite{1}). However, one
might
expect that the very early universe had a non-trivial topology and (or)
that it was
very hot. This renders interesting to investigate the phenomenon of
dynamical chiral symmetry breaking in curved spacetime with non-trivial
topology.

 The present work is specifically devoted to the study of the NJL
model in a curved
spacetime of the form ${\cal M}^3 \times {\rm S}^1$, where
${\rm S}^1$ is the one-dimensional sphere and ${\cal M}^3$  a
three-dimensional, arbitrarily
curved manifold of trivial topology. Here, we will only consider
the
standard choice of boundary conditions (periodic and antiperiodic) for
the fermion on the one-dimensional sphere S$^1$ \cite{8}.     
However, more general choices of the boundary conditions are, of course,
possible \cite{7}, in particular those in which the fermion $\psi$
acquires an extra phase
$\exp{(2 \pi \varphi )}$ with arbitrary $\varphi $ (see the second ref.
in
\cite{7}) each time it goes along ${\rm S}^1 $. The $\varphi =0 $ case
corresponds to periodic boundary conditions and $\varphi = 1/2 $
corresponds to antiperiodic boundary conditions.

 In the next section we calculate the effective potential for the NJL
model in the spacetime ${\cal M}^3 \times {\rm S}^1 $, in the
$1/N$-expansion and for the
linear curvature approximation. In section 3, the phase structure of
the effective potential is discussed and the dynamically generated
fermionic mass is calculated. The existence of topology-- and
curvature--induced phase
transitions from the chiral symmetric phase to the non-symmetric one
is established for different regions of the parameters of the theory.

\section{The effective potential for the NJL model}
    Let us start the calculation of the effective potential for the
NJL model
in the $1/N$-expansion. The classical action of the theory in curved
spacetime is given by
\begin{equation}
S = \int d^4 x \sqrt{g} \left\{ \bar{\psi} i \gamma ^\mu (x)
\nabla _\mu \psi + \frac{\lambda}{2 N} \left[ \left( \bar{\psi} \psi
\right) + \left( \bar{\psi} i \gamma_5 \psi \right)^2 \right] \right\},
\label{1}
\end{equation}
where $N$ is the number of fermions and the rest of the notation is
standard \cite{4,5}.
 It is more useful in actual calculations to work
with the following
equivalent action, in which the auxiliary fields $\sigma $ and $\pi $
are introduced:
\begin{equation}
S = \int d^4 x \sqrt{g} \left[ \bar{\psi} i \gamma ^\mu (x)
\nabla _\mu \psi - \frac{N}{2 \lambda} \left( \sigma ^2 + \pi ^2
\right) - \bar{\psi} \left( \sigma + i \gamma_5 \pi \right) \psi
\right] \ .           \label{2}
\end{equation}
Already in flat spacetime is it known that the global chiral symmetry
---which is the classical symmetry of the theory (\ref{1}) or
(\ref{2})--- is spontaneously broken when the coupling constant
$\lambda$
exceeds some critical value $\lambda_c$. The contribution of the
external gravitational field to this effect has been discussed in ref.
\cite{5}.
    Our purpose here will be to determine the influence of combined
effects, namely of external gravity and non-trivial topology,
simultaneously, on the dynamical chiral symmetry
breaking and restoration. Notice that such a study cannot be done for
the most simple low-dimensional analogue of the NJL model, namely
 the $D=2$
Gross-Neveu model \cite{9}, neither for the ---a little more
complicated--- $D=2$
Schwinger \cite{12} or $D=2$ Thirring  \cite{13} models. In all
those cases
one can study only the isolated effects of either non-trivial topology
(non-zero temperature \cite{10}) or external gravity \cite{11} (see also
\cite{1}) on the chiral symmetry breaking pattern.

Let us now proceed with the explicit calculations. First of all, we
   introduce the semiclassical effective action $S_{eff}$ as
\begin{equation}
Z[0,0] = \int D\sigma \ D\pi \exp \left( i N S_{eff}\right) \ ,
\label{seff definition}
\end{equation}
where $Z$ is defined, in the usual way, to be the generating functional
\begin{equation}
 Z[\eta ,\bar{\eta} ] = \int D\psi \ D\bar{\psi} \ D\sigma
\ D \pi \ \exp \left(iS+i \bar{\eta} \psi + i \bar{\psi} \eta \right) \
{}.
\end{equation}
    This $S_{eff}$ is trivially solved by integrating over the
variables $\psi$ and $\bar{\psi}$, what amounts to a simple
gaussian integration, namely
\begin{equation}
S_{eff} = -\int d^4 x \sqrt{g} \left[ \frac{1}{2 \lambda}
\left(\sigma^2+\pi^2 \right) \right]+i \log \det \left[ i \gamma^\mu
(x) \nabla_\mu- \left( \sigma + i \gamma_5 \pi
\right) \right] \ .         \label{determinant}
\end{equation}
Moreover, $S_{eff}$ is the leading term in the large-$N$
expansion of the effective action $\Gamma$.

    The effective potential $V \left(\sigma,\pi \right)$ is
defined by $V \left(\sigma,\pi \right)= -
\Gamma \left(\sigma,\pi \right)  / (N \cdot {\rm Volume}) $, with
constant configurations for the fields. Thus, to leading order in
the $1/N$-expansion, we have
\begin{equation}
V \left(\sigma , \pi \right) = \frac{1}{2 \lambda} \left(
\sigma^2 + \pi^2 \right) + i\,{\rm Sp}\  \log <x\mid  \left[ i
\gamma^{\mu}
\left( x \right) \nabla_{\mu} - \left(\sigma + i \gamma_5 \pi
\right) \right] \mid x>
\label{sand}
\end{equation}
The second term is directly related to the Green function given
by the equation \[ \left[ i \gamma ^{\mu }(x) \nabla _{\mu } - s
\right] S(x,y) = \delta ^4 (x,y) , \]
where $\delta$ refers to the scalar, coordinate independent Dirac
delta functional for the given manifold.
This fact can be seen by directly applying the operator expression
\[ \log \left( \frac{A-s}{A} \right) = - \int_{0}^{s} d r \frac{1}{A-r}
 \]
to (\ref{sand}).

    To calculate the second term of the lhs of equation (\ref{sand}),
we make use of Schwinger's proper-time method. First, we write
\begin{equation}
V \left(\sigma , \pi \right) = \frac{1}{2 \lambda} \left(
\sigma^2 + \pi^2 \right) - i\,{\rm Sp}\  \log <x\mid \left(i\gamma^{\mu}
\nabla_{\mu} - s \right)^{-1}\mid x>,
\end{equation}
and using the known expression for the propagator of a free Dirac field
in a weakly-varying gravitational background (which has been obtained
in terms of the Riemann normal coordinate expansion \cite{14}), we will
then write the
effective potential with accuracy up to linear curvature terms.
Notice also that a summation over the two
inequivalent spin
structures which are admitted by the spacetime ${\cal M}^3 \times {\rm
S}^1$ will be performed \cite{8}. Of course, one can consider also the
corrections
corresponding to periodic and antiperiodic (non-zero temperature)
boundary conditions independently. We will make a few comments about
this point later.

    Thus, the effective potential is found to be
\begin{eqnarray}
V\left( \sigma , 0 \right) & = & \frac{1}{2 \lambda} \sigma^2
-i\,{\rm tr}\  \displaystyle \int_{0}^{\sigma} d s \frac{1}{L}
\sum_{n=-\infty}^{n=\infty} \sum_{p=0,1}
\int \frac{d^3 k}{\left( 2 \pi \right) ^3} \left[ \left( \gamma^a
k_a + s \right) \frac{1}{k^2-s^2} \right. \nonumber  \\
& & -\frac{1}{12} R \left( \gamma^a
k_a + s \right) \frac{1}{\left( k^2 - s^2 \right)^2} + \frac{2}{3}
R_{\mu \nu} k^{\mu} k^{\nu} \left( \gamma^a k_a +s \right)
\frac{1}{\left( k^2 - s^2 \right) ^3}  \nonumber  \\
& & \left. - \frac{1}{2} \gamma^a {\cal
T}^{c d} R_{cda \mu} k^{\mu} \frac{1}{\left( k^2 - s^2 \right)^ 2}
\right] ,        \label{calcul}
\end{eqnarray}
where one should integrate over \( k^0, k^1, k^2 \) and sum over the
coordinate
$k^3$, which is given by: $k^3 = \left(2 n + \delta_{p,1} \right) \pi
/ L$. In expression (\ref{calcul}), {\rm tr}\  only refers to the spinor
indices.
We have set $\pi = 0$, since there is a rotational symmetry in the
fields $\sigma $ and $\pi $, so that it is enough to discuss the
$\sigma \neq 0$ case for the effective potential only.

 Integration over $s$ is immediate. To perform the momentum
integration, one first makes the Wick rotation ($k^0=ik^4$) and puts
then a cut-off
to regularize the resulting expressions. In our case, we simply restrict
\[\left( k^4\right) ^2+\left( k^1\right) ^2+\left( k^2\right) ^2 \leq
\Lambda^2  \ ,\]
so that our cut-off is different, when compared with the cut-off for the
case of trivial topology \cite{5}.

    From now on, we shall call $V_1$ the contribution to the
effective potential which comes from the logarithm of the determinant
of the operator which appears in equation (\ref{determinant}).
After carrying out the integrations over $s$ and the momenta, we are
led to the following expression for the contribution to $V_1$ coming
from the $p=0$ case, namely, purely periodic boundary conditions
(this corresponds to the contribution to $V(\sigma , 0)$ obtained by
taking only the $p=0$ term in equation (\ref{calcul})). Of course, we
want to study $V_{1}$, which is given by the sum over the two values of
$p$, but ---as we discuss below--- $V_{1}$ may be written down
immediately once the $p=0$ contribution has been worked out. We obtain
\begin{eqnarray}
V_{1}^{p=0} & = & - \frac{1}{L} \displaystyle \sum_{n=- \infty }^{n = \infty }
\left\{ \frac{1}{3 \pi ^2} \,  \Lambda ^3 \log \left(
1+\frac{\sigma ^2}{\Lambda^2 + \frac{4 \pi ^2 n^2}{L^2}}
 \right) \right.   \nonumber \\
&& +  \frac{2}{3\pi^2} \left[ \sigma^2 \Lambda - \left( \frac{4 \pi^2
n^2}{L^2} + \sigma^2 \right) ^{3/2}
\arctan \left( \frac{\Lambda}{\sqrt{\frac{4 \pi^2 n^2}{L^2} +
\sigma^2 }} \right) + \left( \frac{2 \pi n}{L} \right) ^3 \arctan
\left( \frac{\Lambda L}{2 \pi n} \right) \right] \nonumber \\
&& + \frac{R}{3 \left( 2 \pi \right) ^2 } \left[ \frac{2 \pi n}{L}
\arctan \left( \frac{\Lambda L}{2 \pi n} \right) - \sqrt{ \frac{4 \pi^2
n^2}{L^2} +\sigma^2} \ \arctan \left( \frac{\Lambda }{\sqrt{ \frac{4
\pi^2 n^2}{L^2} + \sigma^2 }} \right) \right]  \nonumber \\
&& - \left. \frac{R}{2 \left( 3 \pi \right) ^2 } \left[ \frac{ \left(
\frac{2 \pi n}{L} \right) ^2 \Lambda }{\Lambda^2 + \frac{4 \pi^2
n^2}{L^2} } - \frac{ \left( \frac{4 \pi^2 n^2}{L^2} + \sigma^2 \right)
\Lambda }{\frac{4 \pi^2 n^2}{L^2} + \sigma^2 + \Lambda^2 }
\right] \right\} \ .                       \label{3h}
\end{eqnarray}

    To simplify this expression we use standard techniques drawn from
complex analysis, such as the expression
\begin{equation}
\displaystyle \sum_{n=-\infty }^{\infty} f \left( \frac{2 \pi n }{L} i \right)
=
\frac{L}{2 \pi i} \int_{-i \infty }^{i \infty } dp\, \frac{1}{2} \left[
f(p)+f(-p) \right] + \frac{L}{2 \pi i}
\int_{-i \infty + \epsilon }^{i \infty
+ \epsilon } dp \, \frac{f(p)+f(-p)}{\exp{(Lp)} -1}.      \label{trk}
\end{equation}
One has just to identify the function $f$ for each term in equation
(\ref{3h}) and then perform the integrals in (\ref{trk}).
Another remark is in order here: when computing the term
\[
- \frac{2}{L} \frac{1}{3 \pi^2}  \sum_{n=\infty }^{\infty }
\left[ \sigma^2 \Lambda - \left( \frac{4 \pi^2 n^2}{L^2}
+ \sigma^2 \right) ^{3/2}
\arctan \left( \frac{\Lambda}{\sqrt{\frac{4 \pi^2 n^2}{L^2} +
\sigma^2 }} \right) + \left( \frac{2 \pi n}{L} \right) ^3 \arctan
\left( \frac{\Lambda L}{2 \pi n} \right) \right] ,
\]
it is better to rewrite it as
\begin{eqnarray}
- \lim_{\sigma ' \rightarrow 0 } \frac{2}{L} \frac{1}{3 \pi^2}
\sum_{n=\infty }^{\infty }  &&
\left[  \sigma^2 \Lambda - \left( \frac{4 \pi^2 n^2}{L^2}
+ \sigma^2 \right) ^{3/2} \arctan \left( \frac{\Lambda }{\sqrt{\frac{4
\pi^2 n^2}{L^2} + \sigma^2}} \right) \right.  \nonumber \\
&& - \ \ \ \left. \sigma '^2 \Lambda + \left( \frac{4 \pi^2 n^2}{L^2}
+ \sigma '^2 \right) ^{3/2} \arctan \left( \frac{\Lambda }{\sqrt{\frac{4
\pi^2 n^2}{L^2} + \sigma '^2}} \right)  \right] ,  \nonumber
\end{eqnarray}
since now the expression within square brackets satisfies the properties
which justify the use of equation (\ref{trk}).

    As we said before, once $V_{1}^{p=0}$ has been computed, one may
write immediately
$V_{1}^{p=1}$ and $V_1$, which is given (by definition) by $V_1 =
V_{1}^{p=0} + V_{1}^{p=1}$, because
\begin{equation}
\frac{1}{L} \sum_{n=-\infty }^{\infty } F \left( \frac{2n+1}{L}
\right)  =
\frac{1}{L} \sum_{n=-\infty }^{\infty } F \left( \frac{n}{L}  \right)
-\frac{1}{L} \sum_{n=-\infty }^{\infty } F \left( \frac{2n}{L} \right)
\end{equation}
and, then
\begin{equation}
\frac{1}{L} \sum_{n=-\infty }^{\infty } \left[ F \left( \frac{2n+1}{L}
\right) + F \left( \frac{2n}{L} \right) \right] = \frac{1}{L}
\sum_{n=-\infty}^{\infty } F \left( \frac{n}{L} \right).
\end{equation}
    Now it is apparent that the physics displayed by the model in the
case of purely periodic boundary conditions and in the one where we
consider both spin structures will be essentially the same, since
\[V_{1,L} = V_{1,L}^{p=0} + V_{1,L}^{p=1} = 2 V_{1,2L}^{p=0} .\]
Thus we see that both cases are related by a trivial rescaling of the
length and an overall factor which multipies $V_1$. That is quite an
appealing result.

    However, the last remark does not apply to the case of purely
antiperiodic
boundary conditions, where we only take $p=1$ ---which gives the
themodynamics of a system in three-dimensional space with trivial
topology. In fact, from the last expression we get
\[ V_{1,L}^{p=1}=V_{1,L}-V_{1,L}^{p=0}=V_{1,L}-\frac{1}{2}
V_{1,\frac{L}{2} } .\]
Henceforth, we shall concentrate below only on the analysis of the case
in which
both spin structures are taken into account, as they appear in equation
(\ref{calcul}).

    The first term on the rhs of equation (\ref{trk}) may be
computed without difficulty in all cases. The second term is, in
general, rather
more involved. In order to simplify this contribution as much as
possible in the cases which come from equation (\ref{3h}), one should
pay careful attention to the determination of the integrand along the
contour of integration. After some work, the final result is found to be
\begin{eqnarray}
\frac{V_1}{\Lambda^4}&=&-\frac{2}{3 \pi^2} \left[ 1-\sqrt{1+x^2}+
\frac{1}{l} \log \left(
\frac {\exp{\left( 2l\sqrt{1+x^2}\right) }-1}{\exp{\left( 2l\right) }
-1} \right) \right] \nonumber \\
&& + \frac{1}{6 \pi^2} \left[ \sqrt{1+x^2} - 1 - \frac{3}{2} x^2
\sqrt{1+x^2} + \frac{3}{2} x^4 {\rm arcsinh} \left( \frac{1}{x}
\right) \right] \nonumber \\
&& + \frac{4}{3 \pi^2} \left[ \displaystyle \int_{x}^{\sqrt{1+x^2}} d \tau
\frac{ \left( \tau^2 - x^2 \right) ^{3/2}}{\exp {\left( 2 l \tau
\right) } - 1} - \int_0^1 d \tau \frac{\tau^3 }{\exp{ \left( 2 l
\tau \right) } - 1} \right]  \nonumber \\
&&+\frac{r}{6 \left( 2 \pi \right) ^2}\left[ 1-\sqrt{1+x^2} +
x^2 {\rm arcsinh} \left( \frac{1}{x} \right) \right]   \\
&& + \frac{2 r}{ 3 \left( 2 \pi \right) ^2} \left[ \int_0^1 d
\tau \frac{ \tau }{\exp{\left( 2 l \tau \right) } - 1} -
\int_x^{\sqrt{1+x^2}} d\tau \frac{\sqrt{\tau^2 - x^2}}{\exp{\left(
2 l \tau \right) }-1} \right] \nonumber \\
&& - \frac{r}{2 \left( 3 \pi \right) ^2} \left( 1- \frac{1}{\sqrt{1+x^2}
} \right) + \frac{r}{ \left( 3 \pi \right) ^2} \left[ \frac{1}{
\sqrt{1+x^2}} \frac{1}{\exp{ \left( 2 l \sqrt{1+x^2} \right) }-1} -
\frac{1}{\exp{\left( 2 l \right) }-1} \right]   , \nonumber
\label{exact}
\end{eqnarray}
where $x=\sigma /\Lambda $, $l=L \Lambda $ and $r=R/\Lambda^2 $.

    The value of the field $\sigma $ which satisfies the gap equation
\[ \frac{\partial }{\partial \sigma } V ( \sigma ,0) = 0 \]
gives a dynamical mass to the fermions. This last equation, when written
in
terms of the natural variables $x$, $l$, $r$ and $c$ ($c \equiv \lambda
\Lambda ^2 $), reads
\begin{eqnarray}
0= \frac{V'(x)}{\Lambda ^4}&=& \frac{x^2}{2c} + \frac{5}{6 \pi^2}
\frac{x}{\sqrt{1+x^2}} - \frac{4}{3 \pi^2 } \frac{x}{\sqrt{1+x^2}}
\frac{1}{1-\exp\left( -2l \sqrt{1+x^2}\right) } \nonumber \\
&&+\frac{x}{\pi^2} \left[ x^2 {\rm arcsinh} \left( \frac{1}{x}
\right) -\sqrt{1+x^2} \right] + \frac{1}{2 \pi^2} \frac{x}{\sqrt{1+x^2}}
\nonumber \\
&& \frac{4}{3 \pi^2} \left[ \frac{x}{\sqrt{1+x^2}} \frac{1}{\exp\left(2l
\sqrt{1+x^2}\right) -1} - 3x \displaystyle \int_{x}^{\sqrt{1+x^2}} d \tau
\frac{
\left( \tau ^2 - x^2 \right) ^{1/2}}{\exp(2 l \tau ) -1} \right]  \\
&&+\frac{rx}{3 \left( 2 \pi \right) ^2} \left[ {\rm arcsinh} \left(
\frac{1}{x} \right) - \frac{1}{\sqrt{1+x^2}} \right] - \frac{r}{4 \left(
3 \pi \right) ^2 } \frac{x}{\left( 1+x^2 \right) ^{3/2}}  \nonumber  \\
&&-\frac{2rx}{3\left( 2 \pi \right) ^2} \left[ \frac{1}{\sqrt{1+x^2}}
\frac{1}{\exp\left( 2l\sqrt{1+x^2}\right) -1}-\int_{x}^{\sqrt{1+x^2}} d
\tau \frac{1}{\sqrt{\tau^2 - x^2}}
\frac{1}{\exp(2l\tau )-1} \right] \nonumber \\
&&-\frac{r}{2\left( 3 \pi \right)^2} \left[ \frac{x}{\left( 1+x^2
\right) ^{3/2}} \frac{1}{\exp\left( 2l
\sqrt{1+x^2}\right) -1}-\frac{2l}{1+x^2}
\frac{\exp\left( 2l \sqrt{1+x^2}\right) }{\left( \exp\left(
2l\sqrt{1+x^2}\right) -1 \right)^2} \right]. \nonumber
\end{eqnarray}

It is not possible to produce an exact, analytical expression
for
the dynamically generated fermion mass. Therefore, we will
calculate it below only in a number of limiting cases.

\section{Small-$L$ limit}
    One gets convinced immediately that it is very difficult to
go any further without
doing some simplification. In this section we will consider the
case $
L \Lambda \ll 1$, and we will treat the opposite case $L\Lambda \gg 1 $
in the next one.

    Let us expand $V/\Lambda^4$ in powers of $l$. Assuming
now that $l \sqrt{1+x^2} \ll 1$, we readily obtain
\begin{eqnarray}
\frac{l V(x)}{\Lambda ^4} & = & \frac{x^2}{2g}-\frac{1}{3 \pi^2}
\log{\left( 1+x^2\right) }+\frac{2}{3 \pi^2} \left( -x^2+x^3
{\rm arcsec}  \sqrt{1+\frac{1}{x^2} } \right) \nonumber \\
&&+\frac{r}{3 \left( 2 \pi \right) ^2}\,x\,{\rm arcsec}
\sqrt{1+\frac{1}{x^2} } \ + \frac{r}{
2 \left( 3 \pi \right) ^2} \left( \frac{1}{1+x^2}-1 \right) +
O\left( l^2 \right) ,
\end{eqnarray}
where $g=\lambda \Lambda^2/l$. In order to study the phase
space of the model, one has to compute also the derivative of the
effective potential. That yields
\begin{eqnarray}
\frac{l V'}{\Lambda^4} & = & \frac{x}{g}  - \frac{2}{3 \pi^2}
\frac{x}{1+x^2} + \frac{2}{\pi^2} x^2 {\rm arcsec}
\sqrt{1+ \frac{1}{x^2}} -\frac{4}{3 \pi^2} \frac{x^2}{1+x^2}
\nonumber \\
 && +\frac{r}{3 \left( 2 \pi \right) ^2}{\rm arcsec}
\sqrt{1+ \frac{1}{x^2}} - \frac{r}{\left( 3 \pi \right) ^2}
\frac{x}{\left( 1+x^2 \right) ^2} -\frac{r}{6 \pi ^2} \frac{x}{1+x^2}.
\end{eqnarray}

\subsection{Phase structure in the fixed-curvature case}

    We will here analyze the phase structure of the model in the case
of fixed curvature. In particular, the situations of zero curvature and
constant non-zero curvature will be considered.

\subsubsection{Zero-curvature case}

    As the validity of our results is restricted by the condition
$r \ll 1$, it is worthwhile studying first the case $r=0$. Here it
can be proven that it is impossible to have a first order phase
transition, in fact there is a second order phase transition at
$g_{cr}=\pi^2/2 $. For $g>g_{cr}$ there is chiral symmetry
breaking
and the symmetry is restored for $g<g_{cr}$; in other words, for a
given value of $\lambda$, the symmetry is restored when $l$ grows
beyond a critical point. In the broken phase the  order parameter
is given by $x_{br}=\pi \left( \frac{2}{\pi^2 }- \frac{1}{g} \right) $
or, in terms of the generated mass $m_{gen} =
\frac{2\Lambda}{\pi} - \frac{L \pi}{\lambda}$. It
is straightforward to study the behavior of the value of the effective
potential at $x_{br}$, for varying $g$ or $l$, in this limit of small
compactification length. One finds \[ V(\sigma
_{br}) = - \Lambda^4 \frac{\pi^2}{6l}
 \left( \frac{2}{\pi^2}-\frac{1}{g} \right)^3 \ .\]
    This last expression has the appearance of some kind of
dynamical dimensional reduction
(see ref. \cite{16,17}). In our case we see that the vacuum becomes less
and less energetic as the compactification length shrinks to zero (see
Fig. 1). In this limit we obtain:
\[ \lim _{L \rightarrow 0} \ L V(\sigma _{br}) =-\Lambda ^3
\frac{4}{3 \pi^4} \]


\subsubsection{Influence of the curvature}

    Now we shall study the influence that the presence of curvature
has on this behavior. It is immediate to notice
that for negative values of the curvature, the symmetry is always broken
(the slope of the effective potential is $r/(24 \pi)$ at the
origin, see Fig. 2).
Moreover, for fixed positive values of $r$ (which is kept always small),
as $l$ grows there is a first-order phase transition: it has actually
lost
its continuous character. Now the critical value of the parameter $g$
is \[ g_{cr}= \frac{\pi ^2}{2} \left( 1 + \frac{2 \pi }{3}
\sqrt{\frac{r}{8}} \right) \ . \]
The approximation is consistent with the small-curvature
limit. The difference between the values
of the order parameter $x$ in the broken and disordered phases
at the phase transition is given by $x=
r/8 $ (retaining again only the first correction coming from the
curvature). The influence of curvature on chiral symmetry
breaking in $D=2$ fermionic models is very similar (see refs.
\cite{1,11}).

\subsection{Phase structure in the case of fixed compactification
length}

    We can now, on the other hand, study the situation when $g$ is
held fixed and the curvature takes on different values.

\subsubsection{Case $g<\pi^2/2$}

    If $g<\pi^2/2 $
there is a continuous phase transition at $r=0$ (the symmetry is
broken for negative values of $r$ and is restored for positive
curvature). As $r$ approaches $0$ from below, the order
parameter tends to zero according to the expression \[ x= \frac{\mid
r \mid }{24 \pi \Delta } \ , \]
or $m_{gen} = \frac{\Lambda^{-1} |R|}{24 \pi \Delta}$, being $\Delta
\equiv \frac{1}{g}-\frac{2}{\pi^2}$.
In deriving this result we assume that $\mid r \mid \ll 1$ and that
$\mid r \mid \ll \Delta $.

\subsubsection{Case $g>\pi^2/2$}

    To finish this section, we consider the case $\Delta < 0$ or,
equivalently, $g>\pi^2/2$. To be consistent with the
requirement of small curvature near the critical point, we have to
assume now that $\mid \Delta \mid \ll 1$. In this setting one
may expect that, as $r$ grows, there will be a phase transition from the
ordered to the disordered phase  at some positive
value of the curvature. Retaining again the first terms of the
expansions only, we obtain \[ r_{cr}=\frac{\left( 3 \pi \right) ^2}
{2} \Delta ^2, \] and the value of the order parameter at the
transition is (for the ordered phase)
\[ x= \frac{3 \pi }{4}  \Delta ^2  \ .\]

\section{Large-$L$ limit}

    We shall here consider again the phase structure in the
fixed-curvature case but corresponding now to the opposite limit, when
$L$ is large.

\subsection{Zero-curvature case}
    It is easy to see from equation (\ref{exact}) that, dropping
exponentially vanishing contributions, one may approximate it with
\begin{equation}
\frac{V(x)}{\Lambda^4} = \frac{x^2}{2 c} -\frac{1}{4 \pi^2} \left[
2\left( \sqrt{1+x^2} -1\right) + x^2 \sqrt{1+x^2} - x^4 {\rm
arcsinh}  \frac{1}{x}  \right] ,    \label{largel}
\end{equation}
where c is, as before, $c=\lambda \Lambda ^2$.

    Let us consider the gap equation $V'(x) = 0$. Differentiation
of eq. (\ref{largel}) yields
\begin{equation}
\frac{V'(x)}{\Lambda ^4} = \frac{x}{\pi ^2} \left( \frac{\pi^2}{c} -
\sqrt{1+x^2} + x^2 {\rm arcsinh}  \frac{1}{x}
\right).                     \label{largelder}
\end{equation}
{}From here it is trivial to see that the symmetry is broken when
\[ \lambda > \frac{\pi ^2}{\Lambda ^2} \ . \]
Otherwise, the symmetry is respected by the vacuum.

\subsection{Influence of the curvature}
    If one takes into account the presence of a background gravitational
field, one can check that the terms missing from equations
(\ref{largel}) and (\ref{largelder}) are such that now
\begin{eqnarray}
\frac{V(x)}{\Lambda ^4}&=&\frac{x^2}{2 c} -\frac{1}{4 \pi^2} \left[
2\left( \sqrt{1+x^2} -1\right)  + x^2 \sqrt{1+x^2} - x^4 {\rm
arcsinh} \frac{1}{x} \right]  \nonumber  \\
 && + \frac{r}{6\left( 2 \pi^2 \right) ^2}
\left( 1-\sqrt{1+x^2} + x^2 {\rm arcsinh}  \frac{1}{x}
 \right) -  \frac{r}{2\left( 3 \pi \right) ^2}
 \left( 1 - \frac{1}{\sqrt{1+x^2}}  \right)    \ ,
\end{eqnarray}
and
\begin{eqnarray}
\frac{V'(x)}{\Lambda ^4}&=&\frac{x}{\pi ^2} \left[ \frac{\pi^2}{c} -
 \sqrt{1+x^2} + x^2 {\rm arcsinh} \frac{1}{x}
 \right.                            \nonumber     \\
&&+ \left. \frac{r}{12} \left( {\rm arcsinh} \frac{1}{x}
- \frac{1}{\sqrt{1+x^2}} \right) -  \frac{r}{18\left( 1+x^2
\right) ^{3/2}} \right]   \ .           \label{largelderwithr}
\end{eqnarray}

\subsubsection{Case $\lambda < \pi ^2/\Lambda ^2 $}
    In this situation one may expect that there will be a continuous
phase
transition. After a short calculation, expanding equation
(\ref{largelderwithr}) around the origin and keeping only the two
first leading contributions to $V'(x)$, one sees
that, very near the phase transition, the order parameter is found by
solving
\[0=\frac{\pi^2}{c}- 1-\frac{5}{36}r +\frac{r}{12} {\rm
arcsinh} \frac{1}{x}  + {\cal O}\left( x^2 {\rm arcsinh}
\frac{1}{x} \right) \ . \]
Thus, in a first approximation
\[ x= \left[ \sinh \left( \frac{12}{\mid r \mid }
(\pi^2/c -1 )- \frac{5}{3} \right) \right] ^{-1} \]
(remember that this is valid for negative r). The symmetry is restored
for positive values of r.

\subsubsection{Case $\lambda > \pi ^2/\Lambda^2$}
    Here, a quick analysis of the previous expressions tells us
that in this case there is a first-order phase transition, from the
disordered to the ordered phase, as the curvature grows beyond a
positive
critical value. Notice that this result is very much like the one that
one obtains in the small-$L$ limit (see Fig. 3).

\section{Conclusions}

Recently there has been an increased interest in the NJL model, in
connection with the dynamical symmetry breaking of the electroweak
interactions, using the top quark condensate as order parameter
\cite{3,18}.
    Here, we have investigated the phase
structure of
this model in a curved spacetime with non-trivial topology ---using
the
$1/N$-expansion and working in the linear-curvature approximation---
and we have explicitly shown the possibility of curvature- and (or)
topology-induced phase transitions from a chiral symmetric phase to a
chiral non-symmetric one. In our approximate analysis, we have discussed
the composite field effective potential where summation over two
inequivalent spin structures has been performed. Of course, one can
get
the results corresponding to purely periodic  (or antiperiodic) boundary
conditions as some particular cases of the above study.

    There are different possibilities to extend our analysis. First,
 one can consider different topologies, for example, ${\cal M}^2
\times {\rm T}^2$ or the hyperbolic one ${\cal M}^2 \times {\cal
H}^2/\Gamma $. Second, it would be of interest to study the same
problems treating the external gravitational field exactly. In
particular, one possibility is to take the De Sitter space ${\rm S}_4 $
for
such a calculation. These considerations may draw some connection
with
renormalized quantum gravity in the $1/N$-expansion (see \cite{19}),
that could be certainly important for quantum cosmology near the Planck
scale. We plan to discuss these questions in the near future.

\vspace{5mm}

\noindent{\large \bf Acknowledgments}

SDO would like to thank T. Muta  for helpful discussions and C.
Hill for correspondence. This
work has been partly supported by
CIRIT (Generalitat de Catalunya) and by
DGICYT (Spain), project no. PB90-0022.

\newpage

\noindent{\Large \bf Figure captions}
\bigskip

\noindent{\bf Figure 1}.
Plot of the function $V/\Lambda^4$
for fixed $c=0.05$, $r=0$, and different values of $l$.
\bigskip

\noindent{\bf Figure 2}.
Plot of $V/\Lambda^4$ for fixed $g=5.0404$
(case $g> \pi^2/2$) and for different values of the curvature.
It shows the discontinuous character of the phase transition in
this case.
\bigskip

\noindent{\bf Figure 3}.
Plot of the effective potential in the large-$L$ limit,
corresponding to the case $\lambda >\pi^2/\Lambda^2$
($c$ is held fixed at $c=\pi^2 + 0.15$)
for different values
of $r$. The discontinuous character of the phase transition is clearly
seen.

\end{document}